\documentclass[pra,preprint,aps,superscriptaddress]{revtex4}
\usepackage{amsmath}
\usepackage{graphicx}
\usepackage{tabularx}
\usepackage{amssymb}
\usepackage{natbib}
\usepackage{hyperref}
\usepackage[usenames]{color}
\definecolor{rltred}{rgb}{0.75,0,0}
\definecolor{rltgreen}{rgb}{0,0.5,0}
\hypersetup{colorlinks,linkcolor=rltred,citecolor=rltgreen}

\begin{document}

\title{
Control defeasance by anti-alignment in the excited state.
}

\author{Bo Y. Chang}
\affiliation{School of Chemistry, Seoul National University, Seoul 08826, Republic of Korea}

\author{Seokmin Shin}
\affiliation{School of Chemistry, Seoul National University, Seoul 08826, Republic of Korea}
\email{sshin@snu.ac.kr}

\author{Jes\'us Gonz\'alez-V\'azquez}
\affiliation{Departamento de Qu\'{\i}mica, M\'odulo 13, Universidad Aut\'onoma de
Madrid, 28049 Madrid, Spain}

\author{Fernando Mart\'{\i}n}
\affiliation{Departamento de Qu\'{\i}mica, M\'odulo 13, Universidad Aut\'onoma de
Madrid, 28049 Madrid, Spain}
\affiliation{Instituto Madrile\~no de Estudios Avanzados en Nanociencia (IMDEA-Nanociencia), Cantoblanco, 28049 Madrid, Spain}
\affiliation{Condensed Matter Physics Center (IFIMAC), Universidad Aut\'onoma de Madrid, 28049 Madrid, Spain}

\author{Vladimir S. Malinovsky}
\affiliation{U. S. Army Research Laboratory, Adelphi, Maryland 20783, USA}

\author{Ignacio R. Sola}
\affiliation{Departamento de Qu\'imica F\'isica, Universidad Complutense, 28040 Madrid, Spain}
\email{isola@quim.ucm.es}


\begin{abstract}
We predict {\em anti-alignment} dynamics in the excited state of H$_2^+$
or related homonuclear dimers in the presence of a strong field.
This effect 
is a general indirect outcome of the 
strong transition dipole 
and large polarizabilities typically used to control or 
to induce alignment in the ground state.
In the excited state, however, the polarizabilities have the opposite sign
than in the ground state, generating
a torque that aligns the molecule 
perpendicular to the field, deeming 
any laser-control strategy impossible.
\end{abstract}

\maketitle

\section{Introduction}

Several experiments have reported successful control on the yields
and products of photodissociation
or other photo-induced processes using short and strong laser pulses
\cite{AssionSci98,LevisSci01,DanielSci03,BrixnerPRL04,NuernbergerPNAS10}. 
In principle, strong non-resonant fields can cleanly ({\em i.e.} adiabatically)
manipulate the transition from reactants to products by distorting
the reaction landscape through dynamical Stark-shifts that are sensitive
to the reaction coordinate\cite{SussmanSci06,KimJPCA12,CorralesNat14}.

These manipulations are most effective when the electronic forces 
and gradients are small or comparable to the laser-induced couplings,
allowing to shape true light-induced potentials, commonly referred to as LIPs
\cite{YuanJCP78,BandraukJCP81}.
But the limiting factor is the presence of the coupling, which in 
the dipole approximation is of the form $\mu_{ij}(r) E(t) \cos\theta$,
where $E(t)$ is the laser amplitude, $\mu_{ij}(r)$ the transition dipole, and
$\theta$ the angle between the molecular axis and the field polarization.

From the electronic structure point of view, effective quantum control relies
on finding accessible pairs of states where the transition dipole does not 
decay to zero along the reaction coordinate\cite{SolaAAMO18}. 
This is typically the bottleneck 
of many control scenarios. With limited pulse bandwidth (few lasers operate at 
enough power that most photon wavelengths are within some narrow ranges of 
frequencies)
the wave packet is mostly affected by one or few
transition dipoles to guide its motion, whereas the potential energy
surface is quagmired by conical intersections where the dipole is 
essentially zero\cite{YarkonyRMP96,WorthARPC04,DomckeYarkonyKoppel}. 

In simple molecules ({\em i.e.} diatomic molecules) one can mostly
avoid these problems. Indeed, aligned homonuclear diatomic cations provide
the ultimate examples of highly-controllable molecules\cite{ChangIJQC16}. 
The symmetry-induced charge-transfer at dissociation is responsible for
a huge transition dipole between the pair of states that instead
of decaying asymptotically as the states become degenerate
in the reaction channel, it increases linearly with the internuclear
distance. This dipole can then be used to bind the molecule at any bond 
length, in a process known as
LAMB (laser-adiabatic manipulation of the bond)\cite{ChangPRA03,ChangJCP04,SolaPRA04,JGVJPCA06,ChangPRA10,ChangJCP11}.

There is an additional source acting
against laser control, not due to the vibration but to the rotation
and orientation of the molecule with respect to the field, as
noted in the $\cos\theta$ factor of the interaction. 
Within the LIPs framework, at $\theta = \pi/2$ 
the LIPs show light-induced conical intersections (LICIs) that
often provoke undesired transitions\cite{MoiseyevJPB08,HalaszJPCA12,DemekhinJCP13,HalaszJPCL15}. 
In addition, near this point only 
molecular electronic gradients act on the nuclear wave packet, 
hence dissociative states lose their laser-induced binding properties.
However, it is well known that lasers (and specially strong non-resonant
lasers) force the alignment of the molecular axis with the polarization
axis. Even when no prior alignment is pursued or measured, it is usually 
believed that the concomitant alignment is helping in the control of
the dynamics or, at least, is a resource that the controller may exploit
\cite{BalintKurtiACP08,BrifACP12}.

But while this is the case in the ground state, where one can easily show 
by second order perturbation theory that the static polarizabilities in the 
ground state have to be negative, inducing alignment, it is not necessarily 
so in the excited states. 
In this work we will show that precisely in diatomic homonuclear cations the 
polarizabilities in the excited state are basically just the opposite
of the polarizabilities in the ground state, inducing {\em anti-alignment}.
That is, the effective potential for the rotation in the presence of
the field has a maximum when the molecular axis is aligned with the
transition dipole and a minimum when it is perpendicular to the field.
Then basically all the molecules that are not initially aligned with
the field will dissociate in the excited state, defeating any control
strategy.

The paper relies on a $2$-dimensional model Hamiltonian with soft-core
Coulomb potentials\cite{JavanainenPRA88,SuPRA91,KulanderPRA96}, 
where the nuclear motion is included using the
Ehrenfest approach\cite{Tully1998,TullyJCP12}. 
For the aligned molecules (in a effectively 
$1$-dimensional treatment) the model gives results that are in agreement
with those of full quantum (electron and nuclear) calculations with
the same Hamiltonian\cite{Changarx19}.
But we expect, as explained in the paper, 
the validity of the results to be far more general, as well as 
the process of anti-alignment.

In the remaining of this work we first introduce the model Hamiltonian.
Then we analyze the results obtained in the ground state of molecules
at random orientations
and in the excited state of molecules initially aligned with the
field, before studying the dynamics of molecules initially
misaligned with the field.





\section{Model}

Our goal is to design a simple model that can provide qualitative predictions
of quantum control in one-electron systems, treating strong field laser
couplings with polarized lasers in non-aligned diatomic (or planar) molecules, 
non-adiabatic couplings and ionization
on equal footing. To that end we use soft-core Coulomb potentials in a
$2$-D grid, where the nuclear motion is incorporated in the Ehrenfest
approach. Calling $\tilde{X}_j^{(\alpha)}$ ($j=1,2$ for the $X$, $Y$ coordinates,
$\alpha = 1,2$ for the nuclei of mass $M_\alpha$ and charge $Z_\alpha$) 
the Cartesian coordinates of the nuclei in the plane of motion,
and equivalently $\tilde{x}_j$ the electron coordinates,
collectively written as $\tilde{\bf{X}}$ and $\tilde{\bf{x}}$ 
respectively, and $\psi(\tilde{{\bf x}})$ the electron wave function, 
the electronic degrees of freedom obey the time-dependent Schr\"odinger
equation (in atomic units)
\begin{equation}
i\frac{\partial}{\partial t} \psi(\tilde{\bf x}) = 
-\frac{1}{2} \sum_j \frac{\partial^2} {\partial \tilde{x}_j^2} 
\psi(\tilde{\bf x}) 
+ \left( V_{sc} + V_{int} \right) \psi(\tilde{\bf x}) 
\end{equation}
while for the nuclear degrees of freedom we use the Hellmann-Feynman
force approximation\cite{Tully1998,TullyJCP12},
\begin{equation}
\frac{d^2}{dt^2}\tilde{X}_j^{(\alpha)} = -\frac{1}{M_\alpha} 
\langle \psi(\tilde{\bf x}) \left| 
\frac{\partial (V_{sc} + V_{int} )}{\partial \tilde{X}_j^{(\alpha)}} \right|
\psi(\tilde{\bf x}) \rangle 
\end{equation}
where the soft-core Coulomb potential,
\begin{equation}
V_{sc} = - \sum_\alpha Z_\alpha 
\left[\sum_j \left(\tilde{x}_j - \tilde{X}_j^{(\alpha)} \right)^2 + 
\epsilon^2 \right]^{-\frac{1}{2}}
+ Z_1 Z_2  \left[ \sum_j \left(\tilde{X}_j^{(2)} - \tilde{X}_j^{(1)}\right)^2 \right]^{-1}
\end{equation} 
has analytic derivatives. 
We have chosen $\epsilon = 1/\sqrt{2}$.
In the presence of a linearly polarized external field, 
${\bf E} = {\bf i}E_1  + {\bf j}E_2 $,
in the dipole approximation, the interaction potential is given by
\begin{equation}
V_{int} = \sum_j E_j(t) \left( \tilde{x}_j - Z_2\tilde{X}_j^{(2)} - Z_1\tilde{X}_j^{(1)} \right)
\end{equation}
In this work both nuclei will be protons ($Z_\alpha = 1$,
$M_\alpha = M$). 
It is convenient to decouple the internal and center of mass motion and
work with coordinates relative to the center of mass, not only to reduce the number
of variables, but mainly to avoid having to work with very large grids for the electronic
coordinates, as the whole molecular cation moves in the gradient of the field. 
In order to do so we first disregard the effect of the
magnetic field. 
Calling the nuclear center of mass coordinates ${\bf X}^{CM}$ and defining the distances
with respect to ${\bf X}^{CM}$, ${\bf x} = \tilde{\bf x} -  {\bf X}^{CM}$, 
${\bf X} = \tilde{\bf X} - {\bf X}^{CM}$, we obtain the
same equation for $V_{sc}({\bf x},{\bf X})$ as before [Eq.(4)], 
removing the tilde of the variables. If we neglect the difference between the 
nuclear center of mass and the molecule's center of mass, then
$V_{int} = {\bf E}(t) \left( {\bf x} - {\bf X}^{CM} \right)$.
Within the same level of approximation, we also use the electron mass for the
reduced mass of the system and neglect the mass-polarization term.
Hence, the Ehrenfest equation for $X_j^{(\alpha)}$ remains the same as 
previously [Eq.(3)], but dropping the tilde. 
Now the equations depend 
only on the gradient of $V_{sc}$, while those for ${\bf X}^{CM}$ depend
only on the gradient of $V_{int}$. 
For instance, using polar variables, $X_1^{(1)} = -r/2 \cos\theta$,
$X_2^{(1)} = -r/2 \sin\theta$ ($X_j^{(2)} = -X_j^{(1)}$),
\begin{equation}
V_{sc} = -\left[ (x + \frac{r}{2}\cos\theta)^2 + (y + \frac{r}{2}\sin\theta)^2
+\epsilon^2 \right]^{-1/2} -\left[ (x - \frac{r}{2}\cos\theta)^2 + 
(y - \frac{r}{2}\sin\theta)^2 +\epsilon^2 \right]^{-1/2} + \frac{1}{r}
\end{equation}
$\equiv V_1 + V_2 + r^{-1}$. So,
\begin{equation}
\frac{d^2}{dt^2} r = \frac{1}{M} \left( V_1^3 - V_2^3 \right) \left[
\langle x \rangle \cos\theta + \langle y \rangle \sin\theta \right]
+\frac{r}{2M} \left( V_1^3 + V_2^3 \right) - \frac{2}{Mr^2}
\end{equation}
\begin{equation}
\frac{d^2}{dt^2} \theta = \frac{1}{Mr} \left( V_1^3 - V_2^3 \right)
\left[ \langle y \rangle \cos \theta - \langle x \rangle \sin \theta \right]
\end{equation}
where $\langle x\rangle \equiv\langle \psi({\bf x})| x |\psi({\bf x})\rangle$
and $\langle y\rangle \equiv\langle \psi({\bf x})| y |\psi({\bf x})\rangle$
are the average electronic positions, obtained by integrating the electronic
wave function in the grid.
While for the nuclear center of mass coordinates, 
\begin{equation}
\frac{d^2}{dt^2}{\bf X}^{CM} = \frac{1}{4M} {\bf E}(t)
\end{equation}
which can be immediately integrated.
In the new variables, the TDSE becomes
\begin{equation}
i\frac{\partial}{\partial t} \psi({\bf x}) = 
-\frac{1}{2} \sum_j \frac{\partial^2} {\partial x_j^2} 
\psi({\bf x}) + \left[ V_{sc}({\bf x},r(t),\theta(t))
+ {\bf E}(t){\bf x} - {\bf E}(t){\bf X}^{CM}(t) \right] \psi({\bf x})
\end{equation}
The last term in the parenthesis is a time-dependent scalar that can
be taken away from the TDSE by a unitary transformation of the wave function,
which receives a time-dependent phase due to the accelerated motion of the
center of mass of the charged molecule in the presence of the field,
$\varphi(t) = \int_0^t {\bf E}(t'){\bf X}^{CM}(t') dt'$, where
${\bf X}^{CM}(t)$ is obtained from the solution of Eq.(9).

In our simulations, the initial electron wave function is obtained by 
imaginary time
propagation. To obtain the ground state for a given nuclear arrangement
we initially propagate a 
Gaussian wave function. To obtain the $n$ excited state we start
with a Gaussian times $x^n$ and filter all the lower-energy eigenstates
previously obtained.
For convenience, the molecule is initially aligned in the $x$ axis 
of the grid ($X_2^{(1)} = X_2^{(2)} = 0$) while the laser polarization is 
appropriately rotated by an angle equal to $-\theta$.
The dynamics is solved in a square grid of $256 \times 256$ points 
between $-35$a$_0$ and $35$a$_0$ using the second-order split-operator method.
The kinetic energy operator is evaluated by fast-Fourier transform
and we use imaginary absorbing potentials at the grid edges\cite{MaciasCPL94}.

To represent the behavior of an ensemble of molecules, we typically perform 
$200$ simulations starting with different initial conditions
obtained from random sampling of the Wigner distribution of a Gaussian 
wave packet.
This is either the ground state of the molecule or a wave packet
in the excited state chosen at a particular position 
and with a particular width 
that resembles the
wave packet obtained after ionizing H$_2$ and exciting it with the
third harmonic of an $800$ nm laser \cite{ChangJCP13,ChangJPB15}.
In all the results of this work, we assume that all molecules are
initially non-rotating, that is, the total angular momentum is zero.

\section{Results}

In this work we will analyze 
the orientation, vibration and electron dynamics (the generation of dipoles) in
H$_2^+$ in the presence of a strong {\em static} field.
Although the numerical model assumes some approximations, we expect 
the physical picture regarding the bound states dynamics to emerge
in qualitative agreement with that of a fully quantum treatment.
In particular, the average results obtained from an ensemble of trajectories
(weighted by a Wigner distribution) using the Ehrenfest approximation
for a $1$-D model of H$_2^+$ with a soft-core Coulomb potential 
are very similar to those obtained by solving the quantum $1+1$D 
(electron-nuclear) TDSE, both when we start in a superposition of 
electronic states in the absence of the field, and when we start in the 
excited state (or the dressed electronic state) in the presence of the 
field\cite{Changarx19}.

\begin{figure}
\includegraphics[width=7cm]{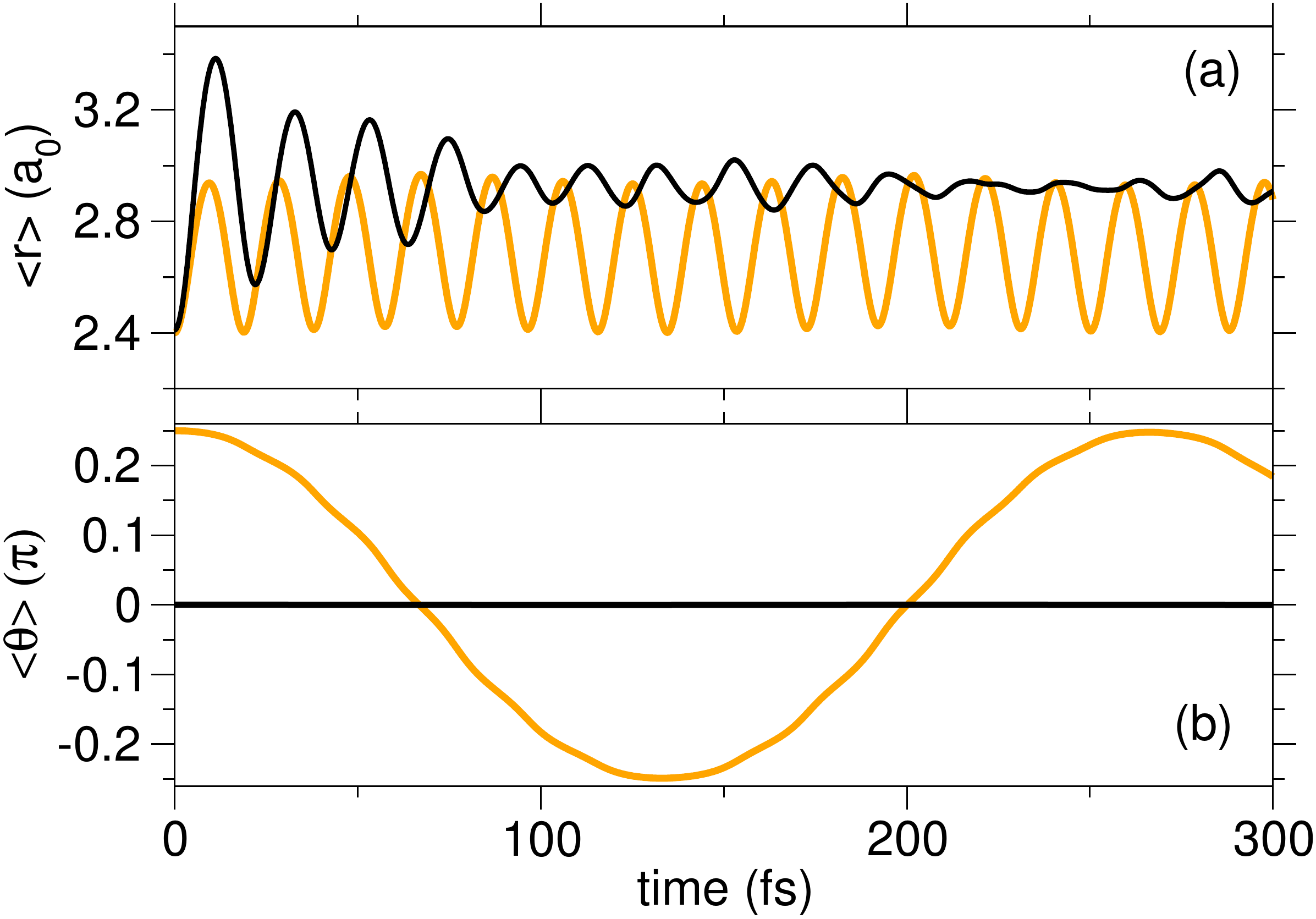}  
\caption{Time evolution of the bond length $\langle r \rangle$ (a) and 
orientation angle $\langle \theta \rangle$
of the molecular axis with respect to the field (b) for an ensemble
of $200$ molecules with random initial orientation and nuclear velocities
and positions according to the Wigner distribution of an initial
Gaussian wave function near the equilibrium distance of the ground state.
The field is constant with $E_0 = 0.02$ a.u. The orange lines shows
a particular trajectory of the ensemble.}
\end{figure}

The $2$-D electronic model has new features: it enables to calculate 
(and observe the influence of) states with orbital angular momentum,
although without the right degeneracy, and more importantly, it allows
to study the role of orientation.
The predictions of the model for the dynamics in the ground state, in
the presence of a strong static field (or a low-frequency continuous-wave laser)
are consistent with the well-known process of alignment.
In Figure 1(a) we show the dynamics of the average bond length 
$\langle r(t) \rangle$ and in Figure 1(b) the average orientation of the 
molecular axis with respect to the field, $\langle\theta(t)\rangle$, 
for a field amplitude of $E_0 = 0.02$ a.u. 
In a particular trajectory (orange line), the molecular axis is 
initially $\pi/4$ degrees with respect to the electric-field polarization 
(in the $x$ axis), while the initial bond length is $2.4$ a$_0$. 
We also show the average results for an ensemble of $200$ molecules
with random orientations with respect to the field and a Wigner distribution
of the internal nuclear coordinates that corresponds to a Gaussian 
wave packet centered at $r_0 = 2.4$ a$_0$ with width $\sigma = 0.5$ a$_0$,
close to the ground state equilibrium bond distance for the $2$-D soft-core
coulomb potential model.
While in a particular trajectory the vibration is almost decoupled from
the rotation in the ground state, given that the internuclear distance
does not vary greatly and the moment of inertia of the molecule remains
fairly constant, the averaged internuclear distances quickly (after the
first oscillations) attenuate the vibrational motion, which is induced
by the anharmonicity of the potential, and hence is sensitive to molecular
orientation. There is also a small additional effect caused by bond-softening 
in the ground state due to the strong field\cite{BucksbaumPRL90,AllendorfPRA91,JolicardPRA92,Giusti-SuzorJPB95}.
A single trajectory exhibits
field-induced alignment, which shows in the pendular dynamics. 
The ensemble of molecules remains uniformly distributed over the
orientations at all times.

\begin{figure}
\includegraphics[width=6cm,angle=-90]{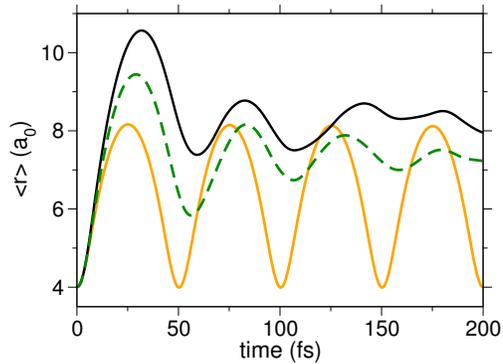}  
\caption{Time evolution of the bond length for an ensemble 
of $200$ molecules initially aligned with the field and in the excited
electronic state. The initial nuclear velocities and positions are
obtained according to the Wigner distribution of a Gaussian wave function 
centered at $\langle r \rangle = 4$ a$_0$ with $\sigma = a_0/2$ 
(black line) and $\sigma = a_0/\sqrt{2}$ (dashed green line).
The field is constant with $E_0 = 0.02$ a.u. The orange lines shows 
a particular trajectory of the ensemble.}
\end{figure}

When the molecule is in the first excited electronic state in the presence
of the field, there is bond-hardening\cite{Giusti-SuzorPRL92,YaoCPL92,ZavriyevPRL93,AubanelPRA93,ChangCPC13}. The electron motion is correlated
to the vibrational motion such that the electron moves away with the
proton against the field, reaching a classical turning point and returning
to the initial position. This was shown in a fully-quantum $(1+1)$-D model
for a molecule aligned with the field\cite{ChangJCP13,ChangJPB15} 
and reproduced with a 1D
Ehrenfest model of the electron moving in the molecular axis\cite{Changarx19}. 
Figure 2 depicts the results using the $2D$-Ehrenfest model for
an ensemble of $200$ molecules initially aligned with a field of
$E_0 = 0.02$ a.u.
The initial nuclear coordinates were obtained from the Wigner distribution
of a Gaussian state centered at $r = 4$ a$_0$, with 
different widths, 
corresponding to the conditions tested in the full quantum
calculation of \cite{ChangJPB15}.
For perfect initial alignment, $\theta(t) = 0$ and the molecules remain
aligned for all times.
As a particular trajectory, we show $r(t)$ 
starting at 
$r(0) = 4$ a$_0$ assuming zero initial vibrational kinetic energy for the 
nuclei.
The average $\langle r(t)\rangle$ reproduces
a decay in the oscillations, due to dephasing.
The amplitude of the
bond dynamics depends on the choice of the initial Wigner distribution.
In our example, the narrower initial distribution ($\sigma = a_0/2$)
reaches larger bond distances. This is because it corresponds to a
broader initial momentum distribution\cite{Changarx19}. For the same reason the
dephasing is somehow stronger for this distribution and the decay
of the oscillation is larger than for the $\sigma = a_0/\sqrt{2}$ case.
In spite of the dephasing, the first oscillations 
are clearly visible marking the regime where the vibrational dynamics
remains coherent for the whole ensemble of molecules, in agreement
with the full quantum calculation.


\begin{figure}
\includegraphics[width=7cm]{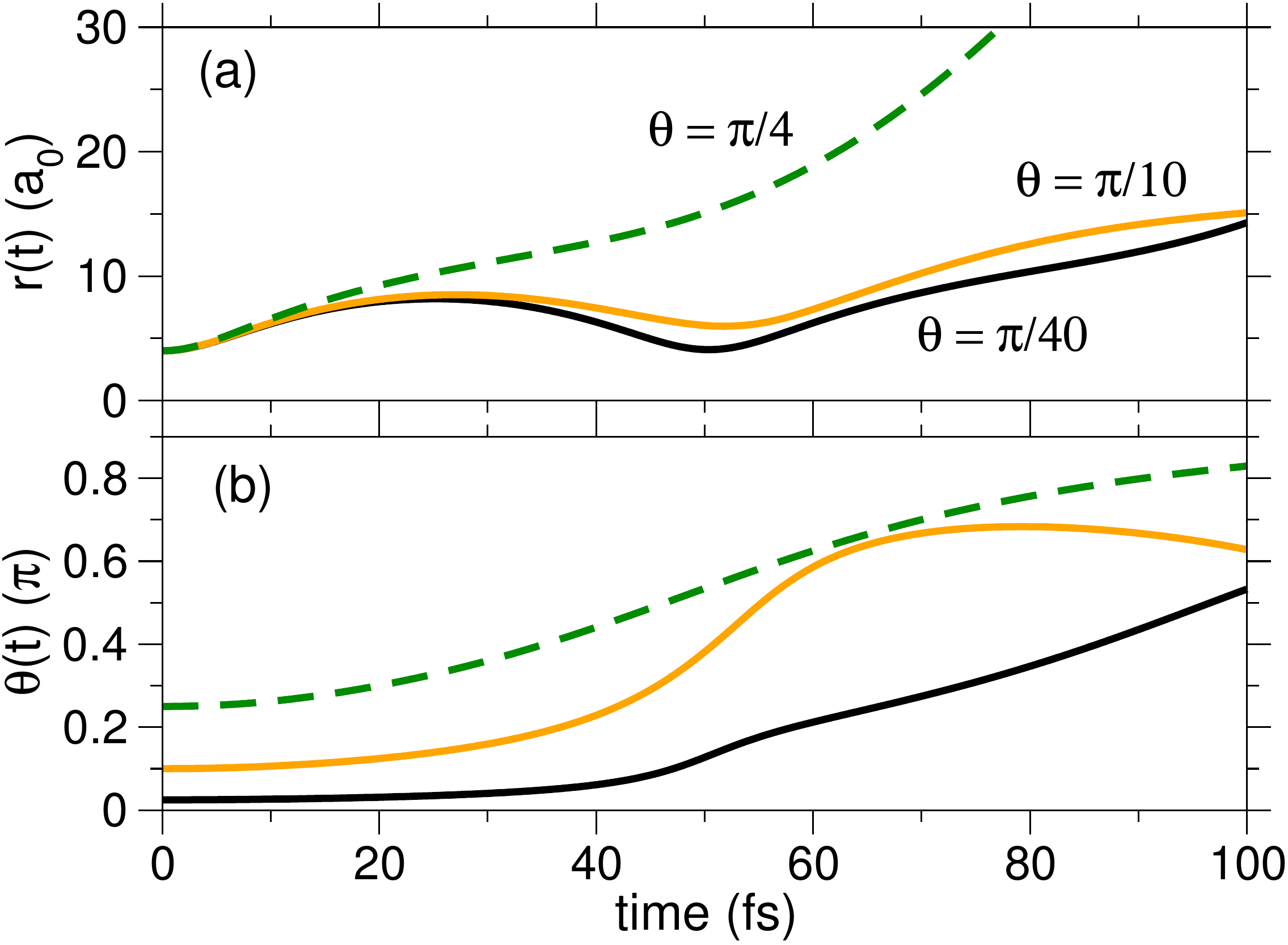}
\caption{Time evolution of the bond length (a) and orientation angle 
of the molecular axis with respect to the field (b) for different
trajectories starting in the excited electronic state, where initially 
$\theta(0) = \pi/2, \pi/10$ and $\pi/40$ $r(0) = 4$ a$_0$ and 
there is no kinetic energy for the internuclear motion.
The field is constant with $E_0 = 0.02$ a.u.}
\end{figure}

\begin{figure}
\includegraphics[width=10cm]{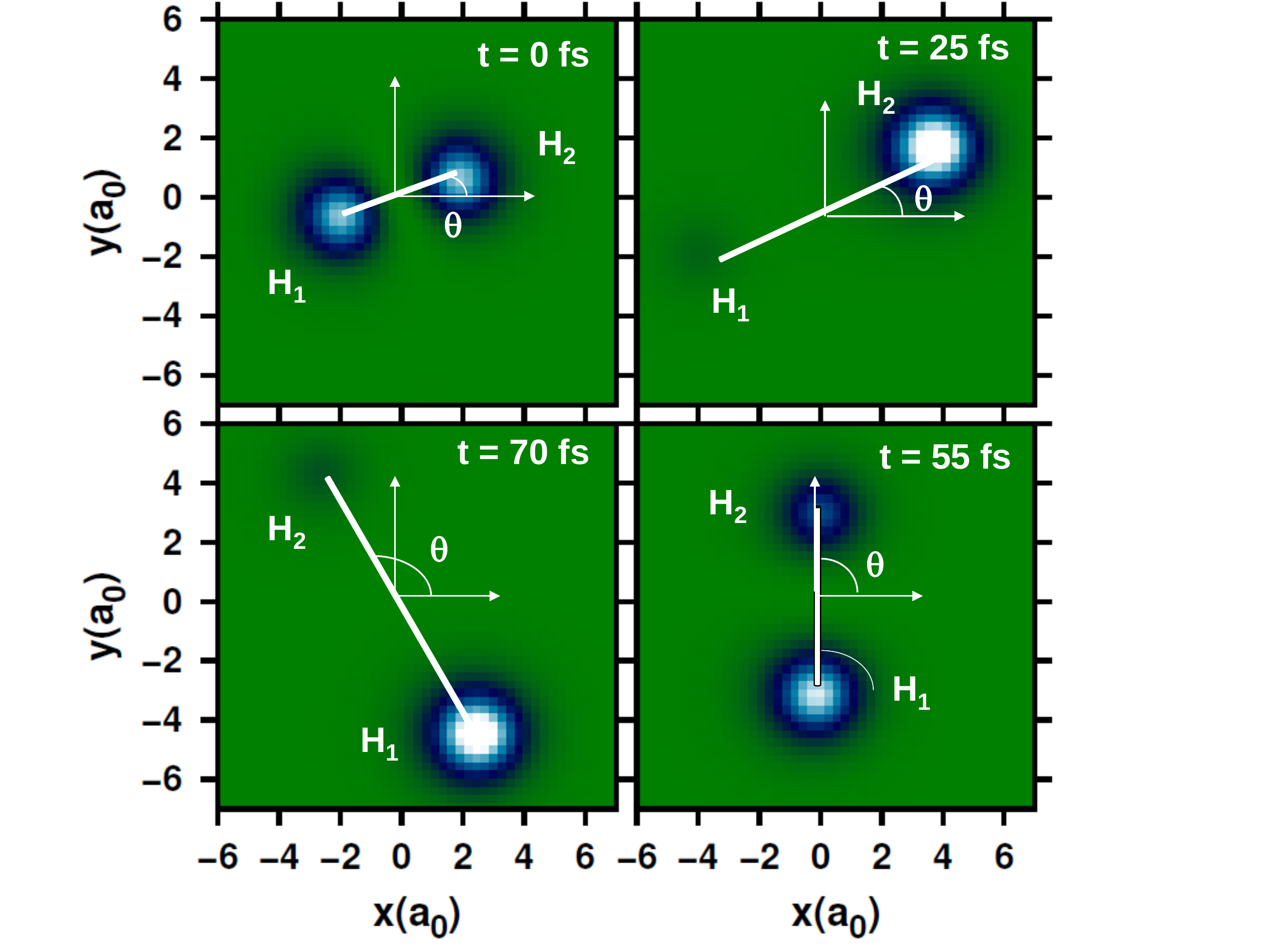}
\caption{Snapshots of the wave packet density at chosen times, where
initially $\theta(0) = \pi/10$, showing the molecular bond as it
enlarges and shrinks before breaking while the molecular axis
anti-aligns with a constant field (in the $x$ axis) of $E_0 = 0.02$ a.u.
The electron density moves with the proton H$_2$ as the bond enlarges, but
then shifts from one proton to the other as $\theta(t)$ crosses $\pi/2$.}
\end{figure}

However, the dynamics differs completely when the molecule is not initially
aligned with the field in the excited state.
In this case, the main driver of the dynamics is the torque against the 
field that is exerted on the molecular axis. This 
{\em anti-alignment} weakens the coupling and eventually leads to 
dissociation when the molecular axis is perpendicular to the field.
In figure 3 we show several examples of the dynamics for different
initial orientations ($\theta(0) = \pi/4, \pi/10$ and $\pi/40$) when the pulse
amplitude is $E_0 = 0.02$ a.u., $\langle r(0) \rangle = 4$a$_0$ and the nuclei
start with zero vibrational kinetic energy in the excited electronic state.

The vibrational period in the excited LIP is approximately
$50$ fs (see Fig.2). While the bond length enlarges initially as expected
(see Fig.3a), the molecule is rotating against the field (see Fig.3b), 
making $\theta$ larger until there is no
coupling at all, at $\theta = \pi/2$, and the molecule dissociates. 
As the coupling with 
the field becomes weaker, the LIP is becoming flatter and the bond recovery
is nonexistent when the molecular axis is initially
very misaligned with the field ({\it e.g.} $\theta(0) = \pi/4$), causing
a fairly direct photodissociation. 
For small initial misalignment ($\theta(0) = \pi/10$ or $\pi/40$)
the coupling is stronger and one can observe some (partial) vibration
in the LIP.
In addition, if the bond is not too stretched when $\theta(t) = \pi/2$
(or is even at minimum internuclear distance, as in the case when
$\theta(0) = \pi/10$), then one observes that
the molecule librates around $\pi/2$.
However, since there is no coupling perpendicular to the field, the electronic 
forces along this component can only contribute to breaking the bond.
In our model this is observed in $X_2^{(2)}(t) - X_2^{(1)}(t)$ 
quadratically increasing with time. 
We have not observed more than two re-crossings of the wave packet back and
forth $\theta = \pi/2$ for any initial conditions, since the dissociation
is relatively fast.

Figure 4 gives hints of the same physical mechanism, also providing
information regarding the role of the electron. We show snapshots of the
wave packet density at different times for the case $\theta(0) = \pi/10$.
Initially the electron density is distributed evenly in both nuclei.
As the bond relaxes, the electron moves with the proton at the right side
($x > 0$), which is the expected result in the excited state 
in the presence of a strong field.
Hence, as $\theta(t)$ crosses $\pi/2$ the electron
density mostly shifts from 
one proton to the other,
where the energy is negative. 
This is because in the excited
state the density is larger at the nuclei sitting at positive values of the
coupling.

\section{Discussion}

In principle, it is not difficult to determine the stereodynamics of
photofragmentation by {\it e.g.} imaging techniques\cite{ChandlerJCP87,EppinkRSI97}.
While very often one observes fragments perpendicular to the bond
axis even in diatomic molecules, they are typically related to 
rotating molecules that give raise to isotropic signals, not to 
polarizability-induced anti-alignment effects.
However, all the results in our simulations imply non-rotating molecules.
The lack of experimental confirmation puts forth the question whether
the results that we observe are an artifact of the approximations in our
model.
In the following discussion we provide compelling reasons to believe that
the results are physical.
As a first approximation, the interaction of the molecule with the
field can be obtained from second order perturbation theory, which
typically is also used to justify alignment. 
For the purpose of understanding the process 
it is enough to analyze
the rotational dynamics, so we can consider a rigid diatomic molecule
in a given electronic state. 
In nonresonant two-photon processes the effective interaction is given by
\cite{StapelfeldtRMP03,SeidemanAAMO05}
\begin{equation}
V_\mathrm{ind} = -\frac{1}{4} E^2(t) \left[ \cos^2\theta \left( 
\alpha_\parallel - \alpha_\perp \right) + \alpha_\perp \right] \ , 
\end{equation}
where $\alpha_\parallel$ and $\alpha_\perp$
are the parallel and perpendicular components of the polarizability
with respect to the molecular axis.
Clearly, the charge transfer resonance responsible
for the strong binding (and large dipole) in H$_2^+$, with transient
dipoles increasing linearly with the internuclear distances and 
electronic state energies approaching asymptotically,
dominates over any other dipole at large bond lengths, hence at
large distances one only needs to account for $\mu_{12} \approx r/2$ 
along the molecular axis \cite{ChangJPB15}
(the dipole between the ground and excited state, or the dipole between
those states related by the charge resonance process).
Then in the ground state, at large bond distance, 
taking\cite{AtkinsMQM}
\begin{equation}
\alpha^{(1)}_\parallel = -2 \sum_{n> 1} \frac{ \langle \psi_1 | \mu_x | 
\psi_n \rangle \langle \psi_n | \mu_x | \psi_1 \rangle }{V_1(r) - V_2(r)}
\approx \frac{r^2}{2\Delta(r)} > 0
\end{equation}
where $\Delta(r) = (V_2(r) - V_1(r))$, is the
detuning, or energy difference between the ground and first excited
electronic states at the chosen internuclear distance, 
and we fixed the molecular axis in the $x$ direction.
Then, $\alpha^{(2)}_\parallel \approx -\alpha^{(1)}_\parallel$ because
of the changing sign in the denominator. Therefore, the same
effect that leads to alignment in the ground state is responsible for
the anti-alignment in the excited state.
On the other hand, because of the large energy gap between $\Sigma$
and $\Pi$ states at large internuclear distances (or the 1s to 2p atomic
energy difference),
$\alpha^{(1)}_\perp \approx \alpha^{(2)}_\perp \approx 0$, 
so the perpendicular component of the polarizability cannot change
the previous prediction. 

In the excited state, the torque will be roughly given by 
$\tau^{(2)} = -dV_\mathrm{ind}/d\theta \approx
\frac{1}{8}r^2 E^2_0 \sin(2\theta) / \Delta(r)$,
so the aligned molecules are in a maximum of the potential, in an unstable
configuration. That is, in the excited state the torque forces alignment
of the molecular axis perpendicular to the field, where the field does
not act and the molecule (in the excited state) dissociates.
Obviously, this rough estimate using second order
perturbation theory fails asymptotically when the electronic states
are degenerate, but the calculation of the LIPs using perturbation
theory for degenerate states shows the same effect.
Note that in our results the calculation is exact (no perturbation
theory is used) but the Hamiltonian is approximate (the electron
is constrained to move in the molecular plane under
a soft-core Coulomb potential).
%


\section{Summary and Conclusions}

We solve the time-dependent Schr\"odinger equation for a $2$-dimensional
soft-core Coulomb potential model of H$_2^+$ in a grid under a strong field,
using the Ehrenfest approach to account for the nuclear motion.
The results for an ensemble of initial conditions reproduce qualitatively
well the known alignment dynamics in the ground electronic state, and
the creation of a huge oscillating dipole in the excited electronic
state, when the molecules are initially aligned with the field.
However, when the molecular axis is misaligned, we observe anti-alignment,
that is, a torque that forces the molecular axis perpendicular to the
field, where the dipole is zero, and the molecule inexorably dissociates.

The results of this work show that laser-induced vibrational trapping
or stabilization in the excited state will be 
impossible in homonuclear cations,
but we expect the presence of anti-alignment dynamics and its effects 
in other laser-induced processes to be very general, and prone
to experimental measurement through {\it e.g.} imaging techniques, whenever
a single transition dipole dominates at large internuclear distances
between two states. 
On the other hand, if the photodissociation is the main goal of the
dynamics, the anti-alignment process avoids laser-induced traps in
the reaction coordinate. In addition, since at $\theta = \pi/2$ there
is a LICI, the anti-alignment may help the wave packet to visit and hence
to probe the dynamics in the vicinity of the LICI.


\section*{Acknowledgments}

This work is supported by the Korean government through the Basic
Science Research program (2017R1A2B1010215) and the National
Creative Research Initiative Grant (NRF-2014R1A3A2030423),
by the Spanish government through the MINECO Project No. CTQ2015-65033-P
and FIS2016-77889-R,
and by the Comunidad de Madrid through project Y2018/NMT-5028.
FM acknowledges support from the ``Severo Ochoa'' Programme for
Centres of Excellence in R\&D (MINECO, Grant SEV-2016-0686) and the
``Mar\'{\i}a de Maeztu'' Programme for Units of Excellence in 
R\&D (MDM-2014-0377).
BYC and IRS thank the Army Research Laboratory for the hospitality in a stay 
during which part of this work was created.

\bibliography{References}
\bibliographystyle{apsrev-nourl}


%
%
%
%

\end{document}